# Skyrmions in Nanotechnology: Fundamental Properties, Experimental Advances, and Emerging Applications

Davi Rodrigues, *Member, IEEE*, Alejandro Riveros, *Member, IEEE*, Andrea Meo, *Member, IEEE*, Emily Darwin, *Member, IEEE*, Vito Puliafito, *Senior Member, IEEE*, Anna Giordano, *Member, IEEE*, Mario Carpentieri, *Senior Member, IEEE*, Riccardo Tomasello, *Senior Member, IEEE,* Giovanni Finocchio, *Senior Member, IEEE*

*Abstract*— **Skyrmions, topologically protected textures, have been observed in different fields of nanotechnology and have emerged as promising candidates for different applications due to their topological stability, low-power operation, and dynamic response to external stimuli. First introduced in particle physics, skyrmions have since been observed in different condensed matter fields, including magnetism, ferroelectricity, photonics, and acoustics. Their unique topological properties enable robust manipulation and detection, paving the way for innovative applications in room temperature sensing, storage, and computing. Recent advances in materials engineering and device integration have demonstrated several strategies for an efficient manipulation of skyrmions, addressing key challenges in their practical implementation. In this review, we summarize the state-of-the-art research on skyrmions across different platforms, highlighting their fundamental properties and characteristics, recent experimental breakthroughs, and technological potential. We present future perspectives and remaining challenges, emphasizing the interdisciplinary impact of skyrmions on nanotechnology.**

This work was supported by the project PRIN2020LWPKH7 "The Italian factory of micromagnetic modelling and spintronics", the project PRIN20222N9A73 "SKYrmion-based magnetic tunnel junction to design a temperature SENSor-SkySens", and the project PRIN2022SAYARY "Metrology for spintronics: A machine learning approach for the reliable determination of the Dzyaloshinskii-Moriya interaction (MetroSpin)", funded by the Italian Ministry of University and Research (MUR). The authors acknowledge the support from the project PE0000021, "Network 4 Energy Sustainable Transition - NEST", funded by the European Union - NextGenerationEU, under the National Recovery and Resilience Plan (NRRP), Mission 4 Component 2 Investment 1.3 - Call for Tender No. 1561 dated 11.10.2022 of the Italian MUR (CUP C93C22005230007). DR and GF acknowledge support from the TOPOCOM project, which is funded by the European Union's Horizon Europe Programme Horizon.1.2 under the Marie Skłodowska-Curie Actions (MSCA), Grant Agreement No. 101119608. ED acknowledges support from the DACH project 200021E_211828 funded by the Swiss National Science Foundation. The authors are with the PETASPIN team and thank the PETASPIN Association (www.petaspin.com). *(Corresponding authors: Davi Rodrigues and Giovanni Finocchio)*

Davi Rodrigues is with the Politecnico di Bari, Bari, 70125 Italia (e-mail: davi.rodrigues@poliba.it).

Alejandro Riveros is with Universidad Central de Chile, Santiago, 8330601, Chile (email: alejandro.riveros@ucentral.cl).

Andrea Meo is with the Politecnico di Bari, Bari, 70125 Italia (e-mail: andrea.meo@poliba.it).

Emily Darwin is with Empa, Swiss Federal Laboratories for Materials Science and Technology, Dübendorf, 8600, Switzerland (e-mail: emily.darwin@empa.ch).

Vito Puliafito is with the Politecnico di Bari, Bari, 70125 Italia (e-mail: vito.puliafito@poliba.it)

Anna Giordano is with the University of Messina, Messina, 98166 Italia (e-mail: anna.giordano@unime.it).

Mario Carpentieri is with the Politecnico di Bari, Bari, 70125 Italia (e-mail: mario.carpentieri@poliba.it)

Riccardo Tomasello is with the Politecnico di Bari, Bari, 70125 Italia (e-mail: riccardo.tomasello@poliba.it).

Giovanni Finocchio is with the University of Messina, Messina, 98166 Italia (e-mail: giovanni.finocchio@unime.it).

Color versions of one or more of the figures in this article are available online at http://ieeexplore.ieee.org

## I. INTRODUCTION

SKYRMIONS were originally introduced to explain the stability of elementary particles, and since their discovery in various physical systems, they have been heralded as a potential breakthrough for nanotechnology [1]. Observed in various physical systems, skyrmions manifest as nanoscale structures with low energy dynamics, exceptional stability, and a wide range of functional responses. The term "skyrmion" derives from physicist Tony Skyrme, who described these structures as topologically defined vortices in a vector field[2], providing an explanation for the stability of fundamental particles. Since then, skyrmions have been predicted and observed in many physical systems, including magnetism, ferroelectrics, photonics, acoustics, and nematic liquids[3], [4], [5], see Figure 1.

The stabilization of skyrmions occurs in physical systems that exhibit long-range order, that can be described by a vector field $\boldsymbol{n}$. In two-dimensional systems, this vector field locally forms a vortex configuration characterized by a well-defined integer topological index known as the skyrmion number. Mathematically the skyrmion number is given by $Q = \frac{1}{4\pi} \int \boldsymbol{n} \cdot (\partial_x \boldsymbol{n} \times \partial_y \boldsymbol{n}) dx dy$, which quantifies the degree of topological wrapping of the long-range order around a unit sphere[2]. Their topologically protected stability makes them robust to continuous deformations, leading to a unique functional response. This stability results in localized bounded modes and particle-like behavior, allowing for exotic physical properties such as emergent chirality and effective dynamical fields.

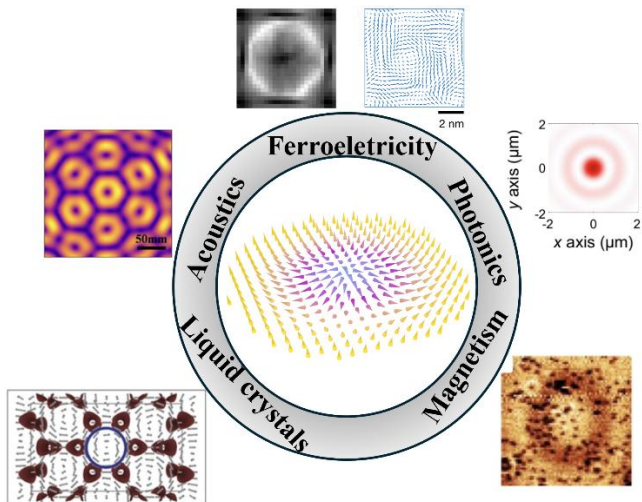

**Figure 1 - Experimental observations of skyrmions in various physical systems.** The central skyrmion configuration has been identified in various media, including[4] ferroelectrics, acoustics[3], liquid crystals[5], optics[6], and magnets[7].

The enhanced stability, low-power driven dynamics and particle-like behavior of skyrmions have inspired many applications in nanotechnology, particularly in sensing, memory and computing. To date, over 5000 publications have explored skyrmions, leading to significant advances in the development of new technologies. Both theoretical and experimental studies have confirmed their existence in various systems, with room temperature stability and long lifetimes [1]. This review aims to summarize these recent discoveries of skyrmions in systems relevant to nanotechnology.

## II. SKYRMIONS IN NANOSYSTEMS

### A. Magnetic systems

Since the prediction and experimental discovery of magnetic skyrmions, they have become a highly active research topic, with an exponentially growing number of publications. In this section, we provide an overview of the history and key properties of magnetic skyrmions, including ferro-, antiferro-, and ferrimagnetic systems. For further reading, we recommend the following review papers and the references therein[1], [8], [9].

Magnetic skyrmions were first proposed theoretically in 1989. The first experimental observation of ferromagnetic Bloch skyrmions in bulk chiral magnets was reported through small-angle neutron scattering in MnSi. Shortly after, direct imaging of skyrmions in thin films using Lorentz transmission electron microscopy (LTEM) confirmed their existence and lattice arrangement. These findings sparked immense interest in skyrmions as potential candidates for next-generation spintronic applications considering the skyrmion itself as the ultimate smallest soliton can be stabilize in magnetic materials.

A defining characteristic of magnetic skyrmions is their electrical manipulation, which makes them attractive for real device implementation and co-integration with mainstream semiconductor technology. It was demonstrated that spin-transfer torque (STT) could drive skyrmions in MnSi. Further studies showed that skyrmions could be efficiently manipulated using electric currents, and their motion in FeGe was visualized via LTEM, providing clear experimental evidence of their dynamic behavior. Theoretical models predicted skyrmion insensitivity to pinning defects and revealed a finite drifting angle with respect to the current direction due to the skyrmion Hall effect.

With the emergence of spin-orbitronics, the interest moved to Néel skyrmions stabilized by interfacial Dzyaloshinskii-Moriya interaction (IDMI) in perpendicular ultrathin ferromagnets in contact with materials, generally heavy metals, with a large spin-orbit coupling. The electrical manipulation of Néel skyrmions through a perpendicular magnetic tunnel junction (MTJ) and spin-orbit torque (SOT) was first theoretically predicted. These skyrmions were later realized experimentally, demonstrating their stability at room temperature and their existence at zero applied field. Current-driven motion of Néel skyrmions was also observed, proving their viability for device applications. However, these experimental observations showed that skyrmions were strongly influenced by pinning defects, contrary to initial predictions. The confirmation of the finite skyrmion Hall angle was demonstrated experimentally. In recent years, stabilization strategies for ferromagnetic skyrmions have also focused on magnetic multilayers with asymmetric interfaces, which allow for stronger thermal stability of skyrmions in the form of tubes, tunable material properties, and reduced skyrmion size.

Beyond electrically driven skyrmion motion, electrical nucleation and detection are also required. Skyrmions can be nucleated from the conversion of domain walls in a constricted geometry, a method that was proposed theoretically and later demonstrated experimentally. Alternatively, geometrical defects can facilitate skyrmion nucleation, as both predicted and observed in experiments. Localized spin-polarized currents allow for precise control over single skyrmion nucleation as well as the engineering of magnetization defects via helium ion or laser irradiation.

The anomalous Hall effect was initially used for skyrmion electrical detection, followed by tunneling magnetoresistance (TMR) detection in MTJs. The relationship between TMR and skyrmion presence was later confirmed using magnetic force microscopy, and more recently by operando magnetic microscopy. An alternative approach to enhancing the skyrmion TMR signal involves combining MTJs with skyrmion-hosting magnetic multilayers[10].

One of the main challenges associated with ferromagnetic skyrmions is the finite skyrmion Hall angle (Magnus effect), which can lead to the annihilation of skyrmions carrying information bits. Antiferromagnetic (AFM) skyrmions have been proposed as a solution since their net magnetization cancels out, allowing for straight-line motion without transverse deflection. Theoretical studies have demonstrated the stability and robustness of AFM skyrmions against external perturbations. However, despite strong theoretical support, experimental observations remain scarce due to challenges in imaging and manipulating antiferromagnetic materials.

A promising alternative to AFM skyrmions is the synthetic antiferromagnetic (SAF) skyrmion. In SAF systems, one non-magnetic material (i.e. Ruthenium, Iridium) is sandwiched between two ultrathin ferromagnetic layers which antiferromagnetically coupled through the Ruderman–Kittel–Kasuya–Yosida (RKKY) interlayer exchange interaction whose sign depends on the thickness of the non-magnetic material. SAF skyrmions were first proposed theoretically, predicting the suppression of the skyrmion Hall effect and enhanced mobility. More recently, SAF skyrmions were observed experimentally[11], and were electrically manipulated, proving their potential for spintronic applications.

Ferrimagnets have also emerged as an attractive platform for skyrmions due to their tunable magnetic properties, allowing ferrimagnetic skyrmions to exhibit lower velocities than AFM and SAF counterparts while maintaining high stability. The first experimental observations of ferrimagnetic skyrmions demonstrated their efficient manipulation via spin-polarized currents. More recent studies have confirmed the potential of ferrimagnetic skyrmions for energy-efficient data storage and processing.

*B. Ferroelectric systems*

Polar skyrmions are non-coplanar swirling field textures in ferroelectric materials, analogous to their well-studied magnetic counterparts, but stabilized by extrinsic effects rather than intrinsic chiral interactions. While magnetic skyrmions arise due to the presence of chiral interactions such as DMI, ferroelectrics lack such chiral stabilization mechanisms. Instead, polar skyrmions arise from the interplay of confined geometries, dipolar interactions, and strain effects, leading to their stabilization in nanoscale ferroelectric structures[12]. These topological textures can manifest as bubble domains-laterally confined, sub-10 nm spheroidal regions with polarization opposite to the surrounding ferroelectric matrix-resulting in a mixed Néel-Bloch character at their domain walls. Such bubble domains, observed in ultrathin $PbZr_{0.2}Ti_{0.8}O_3/SrTiO_3$ heterostructures, undergo topological transitions under electric fields, highlighting their potential for functional applications[12]. Theoretical simulations further show that skyrmion-like topological textures can be induced in conventional ferroelectrics, such as $PbTiO_3$, through controlled Bloch-type domain wall configurations, enabling isotopological and topological transitions driven by external fields and temperature[12].

The experimental realization of room-temperature polar skyrmions in $(PbTiO_3)n/(SrTiO_3)n$ (PTO/STO) superlattices was a major breakthrough, revealing their emergent chirality, negative capacitance, and enhanced dielectric permittivity[13]. These topological structures exhibit a negative permittivity shell, which enhances their dielectric response and enables reversible transitions between skyrmion and uniform ferroelectric states under applied electric fields[13]. Real-space imaging of individual skyrmions has confirmed their electrically tunable dynamics, demonstrating field-driven transformations between skyrmion bubbles, elongated skyrmions, and monodomains[14]. Phase-field simulations have further elucidated the stability and evolution of these structures, revealing field-induced topological transitions that can be exploited for electronic applications[12]. Beyond their tunability with electric fields, skyrmion-based polar textures have been integrated on silicon, demonstrating reversible electric field-driven transitions that modulate resistance, a critical step toward high-density non-volatile memory applications[14].

Recent studies have extended the understanding of polar skyrmions by demonstrating their interaction with optical fields. Twisted light carrying orbital angular momentum has been shown to excite and dynamically manipulate polar skyrmions in ultrathin ferroelectric films, transferring their non-zero winding number and inducing periodic transitions between Bloch and Néel configurations. This interaction introduces a novel route to optically controlled topological transitions, further expanding the scope of polar skyrmion-based functionalities[15]. In addition, the rich topological phase space of polar structures has been explored through phase separation kinetics, revealing a variety of modulated polar states such as dipolar labyrinths, labyrinthine phases, and mixed bimeron-skyrmion states that can be controlled by strain and electric fields[14]. Together, these findings establish polar skyrmions as highly tunable topological objects with promising applications in nanoelectronics, optoelectronics, and topological memory devices, exploiting their unique dielectric, chiral, and topological properties for future technological advances.

*C. Photonic systems*

Recent advances in photonics have shown that confined light fields can lead to the formation of skyrmion-like spin structures. The fundamental mechanism driving the formation of optical skyrmions is the spin–orbit interaction responsible for the coupling and interconversion between spin angular momentum and orbital angular momentum, which are induced by the light polarization and the helical wavefronts, respectively. This coupling can drive the formation of structured optical fields, where the local spin orientation exhibits spatial variations similar to those observed in chiral structures, such as the Néel- and Bloch-type skyrmions[16].

A breakthrough in the experimental realization of optical skyrmions has been achieved manipulating surface plasmon polaritons [16], that are evanescent electromagnetic waves occurring at metal-dielectric interfaces. By exploiting surface plasmon polaritons, both optical field-skyrmion lattices and static optical skyrmions, also known as spin skyrmions, have been observed. Furthermore, recent experimental demonstrations have revealed magnetic plasmon skyrmions, which arise from localized spoof plasmons in space-coiling metastructures.

It is also possible to extend the classical optical skyrmion to higher-order topological quasiparticles[16] such as skyrmioniums and target skyrmions, merons, and three-dimensional structures such as hopfions.

What makes optical skyrmions particularly appealing is their deep-subwavelength resolution which makes possible to scale to dimensions below the diffraction limit, allowing to reach tens of nanometers, and the possibility of resolving and manipulating these structures on the femtosecond scale[17]. These properties can be exploited to achieve super-resolution imaging, high-density optical data storage, and nanophotonic metrology, as well as for potential applications in topologically protected optical computing and information processing[16].

*D. Acoustic systems*

Acoustic waves are traditionally described in terms of scalar pressure fields, and due to their spinless nature they have long been considered to be purely scalar waves. However, the interaction of acoustic waves with other physical phenomena in structured materials can give rise to vectorial properties. In particular, the emergence of acoustic spin waves has been extensively studied, where tailored geometries and wavefront manipulations allow the coupling of acoustics with spin-like properties.

Experimental evidence has confirmed the presence of spin structures in acoustic waves, leading to significant advances in acoustic topological phenomena. Researchers have demonstrated an acoustic topological insulator with a non-zero spin Chern number by engineering nearest-neighbor coupling in phononic graphene[18]. In addition, valley pseudospin states with vortex chirality have been observed in phononic crystals. Multipolar pseudospin states have also been realized in a symmetry-broken metamaterial lattice, demonstrating the intricate control of acoustic wave properties[19].

A major breakthrough established the existence of acoustic spin through the observation of spin angular momentum in free-space airborne acoustics, resulting from the polarization rotation of the particle velocity field[20]. Furthermore, skyrmion textures in the acoustic velocity field have been experimentally observed, including the formation of a tunable Néel-type skyrmion lattice generated by counterpropagating Spoor surface acoustic waves in a hexagonal metasurface[3]. By adjusting the phase of six pairs of loudspeakers, precise control of the skyrmion configuration was achieved. Localized acoustic skyrmions have also been realized in a 3D-printed Archimedean spiral structure, where different skyrmion modes emerge naturally without the need for tailored external excitation[21]. More recently, the robust transfer of skyrmion modes in the acoustic velocity field over long distances has been demonstrated, maintaining the integrity of skyrmion textures within a chain of metastructures with minimal distortion[22].

These works have shown that acoustic waves can generate various topological acoustic textures, including skyrmions, in particular shape-matched metasurfaces. Acoustic skyrmion open new avenues for designing ultra-compact and advanced acoustic vector devices to manipulate acoustic waves with flexible and robust acoustic structures[3], [21].

*E. Liquid crystals*

Skyrmions have been found in liquid crystals, where their unique topology can be studied at room temperature using optical techniques. In highly chiral nematic liquid crystals, a quasi-2D skyrmion lattice can emerge as a thermodynamically stable state under confinement, providing an ideal model system for skyrmion research[5]. Recent studies also show that fractional skyrmions and bimeron strings can be engineered in nematic liquid crystals using topological patterning, leading to dynamic transformations driven by light irradiation[23]. The behavior of skyrmions in liquid crystals has also been explored in the context of non-chiral banana-shaped particles, where excluded-volume interactions alone can stabilize skyrmions through the bend-flexoelectric effect, opening new avenues for self-assembled functional materials[24].

*F. Other physical systems*

Skyrmions, as topologically protected spin textures, appear in a variety of physical media, ranging from quantum Hall systems to chiral magnets, liquid crystals, and multiferroics. In the realm of two-dimensional electron gases under strong magnetic fields, skyrmions manifest as low-energy excitations in the quantum Hall regime at filling factor $v = 1$, where they are described as spin-textured quasiparticles interacting through nonuniform spin rotation[25]. The role of disorder in such systems has been studied, showing that even weak disorder can lead to the presence of skyrmions and antiskyrmions, changing their size and density[26]. Furthermore, at low Zeeman energy, skyrmions become the dominant charged excitations in GaAs heterojunctions, highlighting their stability and correlation-driven origin[27].

In multiferroic and magnetic systems, skyrmions arise from chiral interactions mediated by the Dzyaloshinskii-Moriya interaction (DMI). In bilayer transition metal dichalcogenides, intercalated magnetic atoms allow electric field control of skyrmion chirality, paving the way for energy efficient topological magnetism[28]. Similarly, in ferromagnetic/ferroelectric heterostructures, strain-mediated magnetoelectric coupling enables non-volatile skyrmion creation and control via electric fields, providing a promising platform for spintronic applications[29]. In BiFeO₃ films, uniaxial anisotropy tuning can drive a topological transition, yielding highly stable antiferromagnetic skyrmions that can be manipulated by electric fields and spin torque[30]. These diverse material systems demonstrate the universality of skyrmions in condensed matter physics and their potential for next-generation technological applications.

III. PROMISING APPLICATIONS FOR NANOTECHNOLOGY

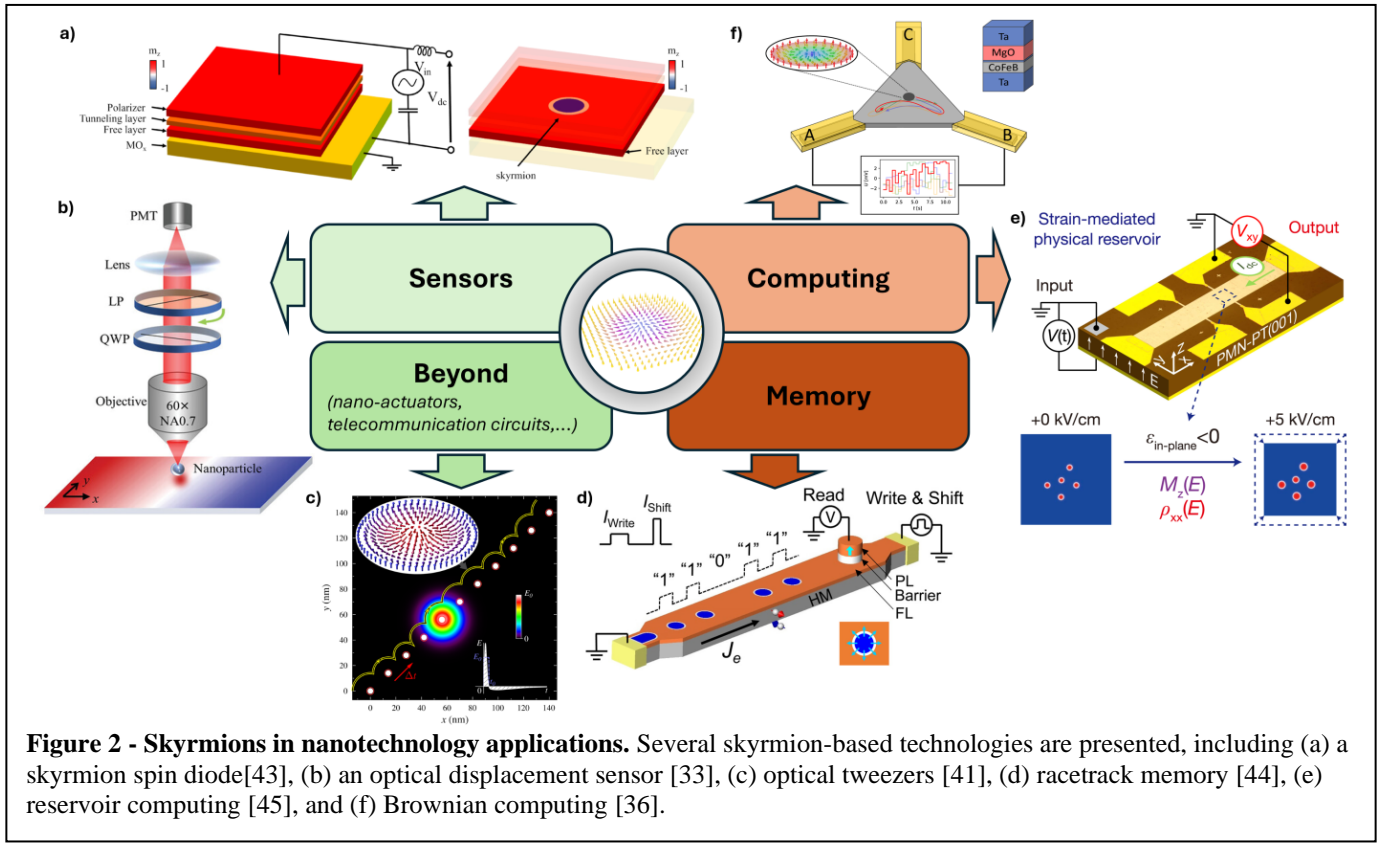

**Figure 2 - Skyrmions in nanotechnology applications.** Several skyrmion-based technologies are presented, including (a) a skyrmion spin diode[43], (b) an optical displacement sensor [33], (c) optical tweezers [41], (d) racetrack memory [44], (e) reservoir computing [45], and (f) Brownian computing [36].

Skyrmions have emerged as promising candidates for various nanotechnology applications due to their unique topological stability, dynamic response to external stimuli, and low-power operation. While the majority of early research focused on magnetic skyrmions due to their initial discovery and widespread interest, recent advances have facilitated knowledge transfer to other fields, leading to innovative applications involving photonic and ferroelectric skyrmions. Figure 2 illustrates key examples of skyrmion-based applications in nanotechnology, highlighting their implementation in photonic and magnetic systems. These applications exploit the intrinsic properties of skyrmions, including their robustness, long lifetime, well-defined low-power excitation modes, quasi-particle behavior, and high mobility, as well as the generation of localized fields. The versatility of skyrmions spans a wide range of technologies, including high-density memory, neuromorphic and probabilistic computing, ultra-sensitive nanoscale sensors, and even next-generation nanoactuators, paving the way for transformative advances in nanotechnology.

## A. Sensors

Skyrmions have shown significant potential for sensing applications in various material platforms, including magnetic, photonic and ferroelectric systems. Their robustness and well-defined excitation modes, which can be triggered with minimal power, make them ideal for highly sensitive and reliable sensing applications. In magnetic systems, skyrmions have been proposed for rotational counting[31], three-dimensional magnetic field detection[32], and microwave signal sensing via spin-torque diode effects that exhibit resonant responses at specific frequencies. Experimental observations reveal that skyrmions allow for a larger selectivity compared to topologically trivial states. Photonic skyrmions can be used for high-precision nanometric displacement detection, optical rulers, and deformation sensing [16], [33], making them suitable for nano-optomechanical applications and precision nanomanipulation. Ferroelectric skyrmions, with their high sensitivity to electric fields and strain, exhibit unique functionalities such as chirality, tunable optical responses, negative capacitance, and giant electromechanical effects. Their compatibility with microelectromechanical systems and integration with flexible 2D materials open new possibilities for next generation nanoelectronic sensors, offering enhanced performance and novel sensing capabilities[34]. The proposed devices highlight the versatility of skyrmions as sensing elements in multiple domains, exploiting their unique topological properties for highly sensitive and reliable detection.

## B. Memory

The topologically stable nature and quasi-particle behavior of skyrmions has sparked considerable interest in their potential use for memory applications, particularly in the development of high-density, non-volatile memory devices. Data can be encoded in skyrmions through various properties, including their density, size, position, or internal characteristics [35]. Their particle-like dynamics make them particularly suitable for applications such as skyrmion-based racetrack memory, in which skyrmions move along a track and stationary write and read components store digital information based on their presence or absence.

In addition, the low-power creation and annihilation of skyrmions by various means further supports their use in memory devices, where their density can be controlled by temperature, current, and electromagnetic fields. Their presence, density, and size can be translated into analog signals, enabling the design of memristive devices[14]. While conventional memristive devices such as resistive random-access memory (RRAM) and phase change memory (PCM) are widely used, skyrmion-based memristive devices offer several advantages, including lower current/voltage consumption, more linear resistance distribution with respect to skyrmion number or size, reduced device-to-device and cycle-to-cycle variations, and improved endurance and retention. However, challenges remain, such as the relatively large bit cell area required to support multilevel states and the smaller on/off ratio compared to binary memory applications.

### C. Unconventional Computing

The use of skyrmions in computing is a promising way to overcome the limitations of conventional hardware by exploiting their nanoscale size, topological stability and energy-efficient dynamics. Their intrinsic properties inspire a growing number of proposals in in-memory computing. Skyrmion-based reservoir computing systems exploit the nonlinear current-voltage characteristics and tunable memory capacity of skyrmions to achieve high-speed, low-power performance in prediction and classification tasks[9]. Thermally-activated skyrmion diffusion also enables Brownian computing, where intrinsic randomness is exploited for stochastic and probabilistic computations, improving fault tolerance and scalability[36]. In addition, skyrmion interactions enable reversible computing through conservative logic gates, supporting low-energy, large-scale Boolean and quantum operations[9]. The quantum properties of skyrmions, including quantized helicity excitations and tunneling effects, open avenues for their integration into quantum computing architectures[37]. Beyond these paradigms, skyrmions are promising candidates for neuromorphic computing, as their accumulation and dissipation mimic synaptic plasticity. This enables the development of artificial synapses with high recognition accuracy in pattern recognition tasks [38]. In these cases, skyrmion accumulation can be leveraged to generate non-binary weights. Furthermore, the electrical manipulation of skyrmion dynamics in different geometries has also been proposed as a means to emulate the leaky-integrate-and-fire behavior[39]. Optical skyrmions open up new possibilities for high-density, noise-resistant photonic computing, exploiting their topological robustness for stable data transmission and arithmetic operations[16]. The thermally excited diffusion of skyrmions has also inspired applications in probabilistic and stochastic computing[36]. These diverse applications highlight the adaptability of skyrmions in unconventional computing, making them strong contenders for future energy-efficient, high-performance computing systems.

### D. Other applications

Beyond their applications in sensors, memory, and computing, skyrmions show remarkable versatility, opening new frontiers in various technological fields. In communications, they have been proposed as data carriers, frequency multipliers, and frequency comb generators, enabling novel signal processing and communication paradigms[6]. Their intrinsic low-power mobility and emergent fields allow skyrmions to mediate interactions with other nanoscale objects, acting as effective nano-actuators. In hybrid superconductor-ferromagnet systems, they facilitate precise control of superconducting vortices and influence dynamic phase transitions[40]. In addition, skyrmions have been proposed as nanoscale tweezers in magnetic, acoustic, and optical systems by exploiting their low-power-driven mobility and localized emergent fields[16], [41]. Furthermore, their compact size and low-energy excitations make them ideal candidates for energy harvesting, further expanding their potential impact in energy efficient technologies[42].

## VII. CONCLUSION AND OUTLOOK

Skyrmions have emerged as a promising avenue for nanotechnology, offering exciting opportunities in magnetism, ferroelectrics, photonics, and acoustics. Their topological protection, nanoscale dimensions, and energy-efficient dynamics position them as strong candidates for next-generation memory, computing, and communication devices. Advances in the stabilization and control of skyrmions, particularly in magnetic and ferroelectric systems, have set the stage for their practical integration in spintronic and electronic applications. In addition, optical and acoustic skyrmions present novel functionalities that could revolutionize wave-based information processing technologies.

Despite these advances, significant challenges remain in achieving full technological realization. Addressing the skyrmion Hall effect, mitigating pinning effects, and refining control mechanisms are critical for reliable skyrmion-based devices. The development of antiferromagnetic skyrmions and hybrid systems, including magnetoelectric coupling in multiferroics, offers promising strategies to overcome these limitations. Similarly, a further understanding of the influence of extrinsic factors such as strain and interfaces in ferroelectric skyrmions will be essential for their robust and reproducible implementation.

Machine learning has emerged as a transformative tool in skyrmion research, aiding in phase classification, magnetic parameter extraction, and predictive modeling of skyrmion dynamics. The use of convolutional neural networks and deep learning techniques has enabled the accurate identification of complex skyrmion textures, even in cases with blurred phase boundaries. In addition, neural ordinary differential equations have shown remarkable efficiency in simulating skyrmion dynamics, providing a powerful alternative to computationally expensive numerical simulations.

With rapid advances in theoretical models, materials engineering, and computational techniques, skyrmion-based nanotechnologies are becoming increasingly feasible. Cross-disciplinary efforts, including the integration of skyrmions with neuromorphic computing, quantum information processing, and AI-driven methodologies, are likely to unlock new functionalities and expand their technological impact. Continued research in this area is expected to drive the development of ultra-compact, energy-efficient, and topologically robust devices, paving the way for the next generation of functional nanotechnologies.

# REFERENCES

[1] G. Finocchio, F. Büttner, R. Tomasello, M. Carpentieri, and M. Kläui, "Magnetic skyrmions: from fundamental to applications," *J. Phys. D. Appl. Phys.*, vol. 49, no. 42, p. 423001, Oct. 2016, doi: 10.1088/0022-3727/49/42/423001.

[2] T. H. R. Skyrme, "A unified field theory of mesons and baryons," *Nucl. Phys.*, vol. 31, pp. 556–569, Mar. 1962, doi: 10.1016/0029-5582(62)90775-7.

[3] H. Ge *et al.*, "Observation of Acoustic Skyrmions," *Phys. Rev. Lett.*, vol. 127, no. 14, p. 144502, Sep. 2021, doi: 10.1103/PhysRevLett.127.144502.

[4] S. Das *et al.*, "Observation of room-temperature polar skyrmions," *Nature*, vol. 568, no. 7752, pp. 368–372, Apr. 2019, doi: 10.1038/s41586-019-1092-8.

[5] J. Fukuda and S. Žumer, "Quasi-two-dimensional Skyrmion lattices in a chiral nematic liquid crystal," *Nat. Commun.*, vol. 2, no. 1, p. 246, Mar. 2011, doi: 10.1038/ncomms1250.

[6] T. He *et al.*, "Optical skyrmions from metafibers with subwavelength features," *Nat. Commun.*, vol. 15, no. 1, p. 10141, Nov. 2024, doi: 10.1038/s41467-024-54207-z.

[7] M. T. Birch *et al.*, "Real-space imaging of confined magnetic skyrmion tubes," *Nat. Commun.*, vol. 11, no. 1, p. 1726, Apr. 2020, doi: 10.1038/s41467-020-15474-8.

[8] A. N. Bogdanov and C. Panagopoulos, "Physical foundations and basic properties of magnetic skyrmions," *Nat. Rev. Phys.*, vol. 2, no. 9, pp. 492–498, Jul. 2020, doi: 10.1038/s42254-020-0203-7.

[9] O. Lee, R. Msiska, M. A. Brems, M. Kläui, H. Kurebayashi, and K. Everschor-Sitte, "Perspective on unconventional computing using magnetic skyrmions," *Appl. Phys. Lett.*, vol. 122, no. 26, Jun. 2023, doi: 10.1063/5.0148469.

[10] Y. Guang *et al.*, "Electrical Detection of Magnetic Skyrmions in a Magnetic Tunnel Junction," *Adv. Electron. Mater.*, vol. 9, no. 1, p. 2200570, Jan. 2023, doi: 10.1002/aelm.202200570.

[11] V. T. Pham *et al.*, "Fast current-induced skyrmion motion in synthetic antiferromagnets," *Science (80-. ).*, vol. 384, no. 6693, pp. 307–312, Apr. 2024, doi: 10.1126/science.add5751.

[12] M. A. Pereira Gonçalves, C. Escorihuela-Sayalero, P. Garca-Fernández, J. Junquera, and J. Íñiguez, "Theoretical guidelines to create and tune electric skyrmion bubbles," *Sci. Adv.*, vol. 5, no. 2, pp. 1–5, Feb. 2019, doi: 10.1126/sciadv.aau7023.

[13] S. Das *et al.*, "Local negative permittivity and topological phase transition in polar skyrmions," *Nat. Mater.*, vol. 20, no. 2, pp. 194–201, Feb. 2021, doi: 10.1038/s41563-020-00818-y.

[14] L. Han *et al.*, "High-density switchable skyrmion-like polar nanodomains integrated on silicon," *Nature*, vol. 603, no. 7899, pp. 63–67, Mar. 2022, doi: 10.1038/s41586-021-04338-w.

[15] L. Gao, S. Prokhorenko, Y. Nahas, and L. Bellaiche, "Dynamical Control of Topology in Polar Skyrmions via Twisted Light," *Phys. Rev. Lett.*, vol. 132, no. 2, p. 026902, Jan. 2024, doi: 10.1103/PhysRevLett.132.026902.

[16] Y. Shen, Q. Zhang, P. Shi, L. Du, X. Yuan, and A. V Zayats, "Optical skyrmions and other topological quasiparticles of light," *Nat. Photonics*, vol. 18, no. 1, pp. 15–25, Jan. 2024, doi: 10.1038/s41566-023-01325-7.

[17] Y. Dai *et al.*, "Plasmonic topological quasiparticle on the nanometre and femtosecond scales," *Nature*, vol. 588, no. 7839, pp. 616–619, Dec. 2020, doi: 10.1038/s41586-020-3030-1.

[18] C. He *et al.*, "Acoustic topological insulator and robust one-way sound transport," *Nat. Phys.*, vol. 12, no. 12, pp. 1124–1129, Dec. 2016, doi: 10.1038/nphys3867.

[19] Z. Zhang, Q. Wei, Y. Cheng, T. Zhang, D. Wu, and X. Liu, "Topological Creation of Acoustic Pseudospin Multipoles in a Flow-Free Symmetry-Broken Metamaterial Lattice," *Phys. Rev. Lett.*, vol. 118, no. 8, p. 084303, Feb. 2017, doi: 10.1103/PhysRevLett.118.084303.

[20] C. Shi *et al.*, "Observation of acoustic spin," *Natl. Sci. Rev.*, vol. 6, no. 4, pp. 707–712, Jul. 2019, doi: 10.1093/nsr/nwz059.

[21] P. Hu *et al.*, "Observation of localized acoustic skyrmions," *Appl. Phys. Lett.*, vol. 122, no. 2, Jan. 2023, doi: 10.1063/5.0131777.

[22] W. Sun, N. Zhou, W. Chen, Z. Sheng, and H. Wu, "Acoustic Skyrmionic Mode Coupling and Transferring in a Chain of Subwavelength Metastructures," *Adv. Sci.*, Jul. 2024, doi: 10.1002/advs.202401370.

[23] Z. Asilehan *et al.*, "Light-driven dancing of nematic colloids in fractional skyrmions and bimerons," *Nat. Commun.*, vol. 16, no. 1, p. 1148, Jan. 2025, doi: 10.1038/s41467-025-56263-5.

[24] R. Subert, G. Campos-Villalobos, and M. Dijkstra, "Achiral hard bananas assemble double-twist skyrmions and blue phases," *Nat. Commun.*, vol. 15, no. 1, p. 6780, Aug. 2024, doi: 10.1038/s41467-024-50935-4.

[25] Y. A. Bychkov, T. Maniv, and I. D. Vagner, "Charged Skyrmions: A condensate of spin excitons in a two-dimensional electron gas," *Phys. Rev. B*, vol. 53, no. 15, pp. 10148–10153, Apr. 1996, doi: 10.1103/PhysRevB.53.10148.

[26] A. J. Nederveen and Y. V. Nazarov, "Skyrmions in Disordered Heterostructures," *Phys. Rev. Lett.*, vol. 82, no. 2, pp. 406–409, Jan. 1999, doi: 10.1103/PhysRevLett.82.406.

[27] S. L. Sondhi, A. Karlhede, S. A. Kivelson, and E. H. Rezayi, "Skyrmions and the crossover from the integer to fractional quantum Hall effect at small Zeeman energies," *Phys. Rev. B*, vol. 47, no. 24, pp. 16419–16426, Jun. 1993, doi: 10.1103/PhysRevB.47.16419.

[28] Z. Shao *et al.*, "Multiferroic materials based on transition-metal dichalcogenides: Potential platform for reversible control of Dzyaloshinskii-Moriya interaction and skyrmion via electric field," *Phys. Rev. B*, vol. 105, no. 17, p. 174404, May 2022, doi: 10.1103/PhysRevB.105.174404.

[29] Y. Ba *et al.*, "Electric-field control of skyrmions in multiferroic heterostructure via magnetoelectric coupling," *Nat. Commun.*, vol. 12, no. 1, p. 322, Jan. 2021, doi: 10.1038/s41467-020-20528-y.

[30] A. Chaudron *et al.*, "Electric-field-induced multiferroic topological solitons," *Nat. Mater.*, vol. 23, no. 7, pp. 905–911, Jul. 2024, doi: 10.1038/s41563-024-01890-4.

[31] K. Leutner *et al.*, "Skyrmion automotion and readout in confined counter-sensor device geometries," *Phys. Rev. Appl.*, vol. 20, no. 6, p. 064021, Dec. 2023, doi: 10.1103/PhysRevApplied.20.064021.

[32] S. Koraltan *et al.*, "Skyrmionic device for three dimensional magnetic field sensing enabled by spin-orbit torques," p. arxiv:2403.16725, Mar. 2024, [Online]. Available: http://arxiv.org/abs/2403.16725

[33] A. Yang *et al.*, "Spin-Manipulated Photonic Skyrmion-Pair for Pico-Metric Displacement Sensing," *Adv. Sci.*, vol. 10, no. 12, Apr. 2023, doi: 10.1002/advs.202205249.

[34] V. Govinden *et al.*, "Spherical ferroelectric solitons," *Nat. Mater.*, vol. 22, no. 5, pp. 553–561, May 2023, doi: 10.1038/s41563-023-01527-y.

[35] S. Luo and L. You, "Skyrmion devices for memory and logic applications," *APL Mater.*, vol. 9, no. 5, 2021, doi: 10.1063/5.0042917.

[36] G. Beneke *et al.*, "Gesture recognition with Brownian reservoir computing using geometrically confined skyrmion dynamics," *Nat. Commun.*, vol. 15, no. 1, p. 8103, Sep. 2024, doi: 10.1038/s41467-024-52345-y.

[37] C. Psaroudaki and C. Panagopoulos, "Skyrmion Qubits: A New Class of Quantum Logic Elements Based on Nanoscale Magnetization," *Phys. Rev. Lett.*, vol. 127, no. 6, p. 67201, 2021, doi: 10.1103/PhysRevLett.127.067201.

[38] T. da Câmara Santa Clara Gomes *et al.*, "Neuromorphic weighted sums with magnetic skyrmions," *Nat. Electron.*, Jan. 2025, doi: 10.1038/s41928-024-01303-z.

[39] J. Liang, Q. Cui, and H. Yang, "Electrically switchable Rashba-type Dzyaloshinskii-Moriya interaction and skyrmion in two-dimensional magnetoelectric multiferroics," *Phys. Rev. B*, vol. 102, no. 22, p. 220409, Dec. 2020, doi: 10.1103/PhysRevB.102.220409.

[40] J. Nothhelfer *et al.*, "Steering Majorana braiding via skyrmion-vortex pairs: A scalable platform," *Phys. Rev. B*, vol. 105, no. 22, p. 224509, Jun. 2022, doi: 10.1103/PhysRevB.105.224509.

[41] X.-G. Wang *et al.*, "The optical tweezer of skyrmions," *npj Comput. Mater.*, vol. 6, no. 1, p. 140, Sep. 2020, doi: 10.1038/s41524-020-00402-7.

[42] D. Kechrakos *et al.*, "Skyrmions in nanorings: A versatile platform for skyrmionics," *Phys. Rev. Appl.*, vol. 20, no. 4, p. 044039, Oct. 2023, doi: 10.1103/PhysRevApplied.20.044039.

[43] D. R. Rodrigues, R. Tomasello, G. Siracusano, M. Carpentieri, and G. Finocchio, "Ultra-sensitive voltage-controlled skyrmion-based spintronic diode," *Nanotechnology*, vol. 34, no. 37, p. 375202, Sep. 2023, doi: 10.1088/1361-6528/acdad6.

[44] G. Yu *et al.*, "Room-Temperature Skyrmion Shift Device for Memory Application," *Nano Lett.*, vol. 17, no. 1, pp. 261–268, Jan. 2017, doi: 10.1021/acs.nanolett.6b04010.

[45] Y. Sun *et al.*, "Experimental demonstration of a skyrmion-enhanced strain-mediated physical reservoir computing system," *Nat. Commun.*, vol. 14, no. 1, p. 3434, Jun. 2023, doi: 10.1038/s41467-023-39207-9.

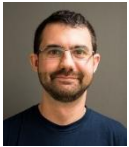

**Davi Rodrigues** (M'20) became a Member (M) of IEEE in 2020. He received the B.Sc. degree in physics from the University of Brasilia, Brasilia, Brazil in 2010. He received the M.Sc. degree in physics from the State University of Sao Paulo, Sao Paulo, Brazil in 2012. He received the Ph.D. degree in applied physics in 2018 from the Texas A&M University, College Station, USA.

He is currently a Jr. Assistant Professor (RTD-a) at the Department of Electrical and Information Engineering of the Politecnico di Bari, Bari, Italy. He was a post-doctoral fellow at the University of Mainz, Mainz, Germany, from 2018 to 2020, and at University of Duisburg-Essen, Duisburg, Germany, in 2020. He was also the scientific coordinator at the JGU Research Center for Algorithmic Emergent Intelligence (https://emergent-ai.uni-mainz.de/) in 2019 and 2020. His main research activity has been the theoretical study and micromagnetic modeling of spintronic devices (spin-torque nano-oscillators, spin-transfer-torque magnetic random-access memory) and their applications to unconventional computing.

Dr. Rodrigues is member of the IEEE Nanotechnology Council Italy-Chapter and of the IEEE Magnetic Society Italy-Chapter.

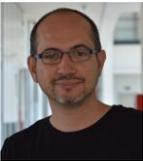

**Alejandro Riveros** (M'21) became a Member (M) of IEEE in 2021. He received the B.Sc. degree in physics and the Ph. D degree in physics from the Department of Physics of the Universidad de Santiago de Chile, Santiago, Chile, in 2008 and 2015, respectively.

He is currently an assistant professor in the Faculty of Engineering and Architecture of the Universidad Central de Chile, Santiago Chile. He was a post-doctoral Fondecyt grant in the Universidad Santiago de Chile, 2018-2019. Currently, his research focuses on the theoretical study and modeling of spintronic devices based on magnetic skyrmions and vortex textures.

Dr. Riveros is member of the IEEE Magnetic society Chile-Chapter and ALMA (Latin American Magnetism Association).

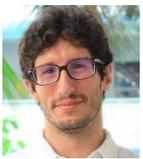

**Andrea Meo** (M'16) became a Member (M) of IEEE in 2016. He received the B.Sc. degree in physics and the M.Sc. degree in physics from the Department of Physics of the University of Milano-Bicocca, Milan, Italy, in 2012 and 2014, respectively. He received the Ph.D degree in physics in from the Department of Physics of the University of York, York, UK, in 2019 working in the computational magnetism group of the University of York.

He is currently an assistant professor (RTT) in the Department of Electrical and Information Engineering of the Politecnico di Bari, Bari, Italy. He was a post-doctoral fellow at the Department of Physics of the University of Mahasarakham, Mahasarakham, Thailand, from 2019 until 2021, and at the Department of Physics of the University of York, York, UK in 2022, and at the Department of Electrical and Information Engineering of the Politecnico di Bari, Italy, in 2022 and 2023. His main research interests include spintronics, magnetic tunnel junctions, magnetic random access memories, antiferromagnets, spin waves, magnetic recording, micromagnetic and atomistic spin modelling.

Dr. Meo is member of the IEEE Nanotechnology Council Italy-Chapter where he serves as Media Manager, and he is also member of the Italian Association of Magnetism (AIM).

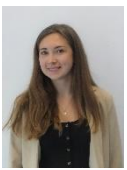

**Emily Darwin** (M' 20) became a Member (M) of IEEE in 2020. She received the MPhys. degree and the Ph.D. degree in physics from the Department of Physics and Astronomy at the University of Leeds, Leeds, United Kingdom, in 2018 and 2024, respectively.

She is currently a postdoctoral researcher at in the Department of Magnetic and Functional Thin Films at Empa, Swiss Federal Laboratories for Materials Science and Technology, Dübendorf, Switzerland. She was a post-doctoral researcher in the Department of Electrical and Information Engineering at the Politecnico di Bari, Bari, Italy, in 2024. Her main research activities include experimental study of thin magnetic films, in particular ferrimagnets and synthetic antiferromagnets, and advancements in magnetic force microscopy.

Dr. Darwin is member of the IEEE Nanotechnology Council Italy-Chapter and of the IEEE Magnetic Society Italy-Chapter.

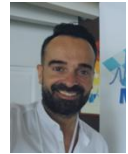

**Vito Puliafito** (M'09-SM'24) became a Member (M) of IEEE in 2009 and a Senior Member (SM) in 2024. He received the M.S. degree in Electronic Engineering and the PhD in Electromagnetic Modeling at the University of Messina, Italy, in 2007 and 2011, respectively.

He is currently a Full Professor of Electrical Engineering with the Department of Electrical and Information Engineering, Politecnico di Bari, Italy. During 2009-2011, he was a visiting PhD student at the University of Salamanca (Spain) and, later, a visiting researcher at the Bogazici University, Istanbul (Turkey), and at FORTH Hellas in Crete (Greece). From 2019 to 2021 he was Assistant Professor at the University of Messina. Since 2021 he has been with the Department of Electrical and Information Engineering, Politecnico di Bari, where he was Associate Professor and became a Full Professor in 2024.

His research interests include modelling magnetic materials at nanoscale, spintronic applications, micromagnetics, and novel approaches for computing. He has published more than 50 papers in international journals and was invited to several international conferences.

Prof. Puliafito is Chair of the IEEE Magnetics Society Young Professionals affinity group and Chair of the IEEE Magnetics Italy Chapter. He is also a co-founder and member of the IEEE Nanotechnology Council Technical Committee on Quantum, Neuromorphic and Unconventional Computing, and a member of the Editorial Board of IEEE Magnetics Letters.

He served in the technical and organizing committees of numerous international conferences and schools, including Director of the 2023 IEEE Magnetics Summer School, General Chair of the 2025 IEEE Conference on Advances in Magnetics (IEEE-AIM 2025), and Program Chair of the 2025 IEEE Conference on Nanotechnology (IEEE-NANO 2025).

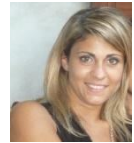

**Anna Giordano** (M'13) became a Member (M) of IEEE in 2013. She received the Ph.D. degree in advanced technologies in optoelectronic, photonic and electromagnetic modeling from the University of Messina, Messina, Italy, in 2014. She is currently an associate professor at the Department of Engineering of the University of Messina, Messina, Italy. Her research focuses on the micromagnetic modelling of spintronic devices, including spin-torque nano-oscillators and spin-torque diodes. She is responsible for the implementation of the code for PETASPIN micromagnetic simulations (a generalization of the GPMagnet code https://www.goparallel.com/index.php/gp-software). PETASPIN integrates unique features; she would like to highlight the possibility of simulating hybrid structures of coupled ferromagnets and ferrimagnets. Her main expertise includes the implementation of numerical methods, modeling, and the micromagnetic design of spintronic devices.

Prof. Anna Giordano is Vice-Chair of the IEEE Magnetic Society – Italy Chapter. She is also she is a member of the Editorial board and Associate Editor of the *IEEE Transactions on Magnetics*.

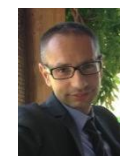

**Mario Carpentieri** (M'11–SM'14) became a Member (M) of IEEE in 2011 and a Senior Member (SM) in 2014. He received the M.S. degree in electronic engineering and the Ph.D. degree in advanced technologies for the optoelectronic and photonic and electromagnetic modeling from the University of Messina, Messina, Italy, in 1999 and 2004, respectively.

He is currently a full professor at the Department of Electrical and Information Engineering of the Politecnico di Bari, Bari, Italy. During 2003–2005, he was a Visiting Researcher in the Department of Applied Physics, University of Salamanca, Spain. From 2005 to 2011, he was an Assistant Researcher at the University of Perugia and University of Calabria, Italy.

Since 2012, he has been with the Department of Electrical and Information Engineering, Politecnico di Bari, where he was an Assistant Professor, became an Associate Professor in 2015, and full professor in 2019. His current research interests include micromagnetic modeling of a variety of spintronic nanostructured materials and devices, including microwave nano-oscillators and diodes based on the spin-torque and spin-orbit effects.

Prof. Carpentieri is Chair of the IEEE Nanotechnology Council Italy-Chapter, and a member of the IEEE Magnetic Society. He is currently a member of the Editorial board and Associate Editor of the IEEE Transactions on Magnetics and Associate Editor of Scientific Reports (Nature). He is co-inventor of 2 patents, and co-authors of more than 130 articles published in well-established international journals (Phys. Rev. Lett., Nat. Comm., Nat. Elec., Appl. Phys. Lett., etc). He is co-founders of one start-up company for the development of parallel computation. He served on many technical program committees of international conferences and organized two international conferences as general Chair and he has been a member of several Program Committees.

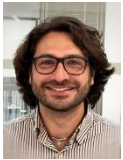

**Riccardo Tomasello** (M'14–SM'23) became a Member (M) of IEEE in 2014 and a Senior Member (SM) in 2023. He received the B.Sc. degree in industrial engineering and a M.Sc. degree in material science and Engineering from the University of Messina, Messina, Italy, in 2010 and 2012, respectively. He received the European Ph.D. degree in system and computer engineering in 2016 from the University of Calabria, Cosenza, Italy.

He is currently an associate professor at the Department of Electrical and Information Engineering of the Politecnico di Bari, Bari, Italy. He was a post-doctoral fellow at the University of Perugia, Perugia, Italy, in 2016 and 2017, and at the Foundation for Research and Technology - Hellas, Heraklion, Greece from 2018 to 2021, where he was also the scientific coordinator of the project "ThunderSKY" (http://thundersky.iacm.forth.gr). His main research activity includes the theoretical study and micromagnetic modeling of spintronic devices (spin-torque nano-oscillators, spin-transfer-torque magnetic random-access memory, microwave detectors, energy harvesters), with particular focus on the micromagnetic analysis of the static and dynamic properties of skyrmions.

Prof. Tomasello is member of the IEEE Nanotechnology Council Italy-Chapter, where he serves as Treasurer, and a member of the IEEE Magnetic Society. He received the *Best Poster Award* at the 61st Annual Conference on Magnetism and Magnetic Materials (2016), New Orleans, USA, and the *Young Researcher Award* at the 2nd IEEE Conference on Advance in Magnetics (2018), La Thuile, Italy, and the *Young Researcher Award* (2019) from the IEEE Magnetic Society Italy-Chapter. He has been a visiting scholar at the Northwestern University, Evanston, Illinois, USA, University of California, Irvine, USA, University of Salamanca, Spain, and Bogazici University, Turkey.

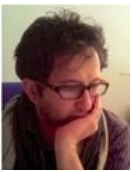

**Giovanni Finocchio** (M'03–SM'12) became a Member (M) of IEEE in 2014 and a Senior Member (SM) in 2023. He received the Ph.D. degree in advanced technologies in optoelectronic, photonic and electromagnetic modeling from the University of Messina, Messina, Italy, in 2005.

He is currently a Full Professor with the Department of Mathematical and Computer Sciences, Physical Sciences and Earth Sciences of the University of Messina and the Director of the PETASPIN laboratory (Petascale computing and Spintronics). His main research interests include spintronics, skyrmions, and unconventional computing.

Prof. Finocchio is member of the Administrative Committee of the IEEE Magnetic Society, and he is member of the IEEE Nanotechnology Council. In the last ten years, he has been on many technical program committees of international conferences and organized more than ten international conferences and workshops as the Chair, Program Committee Member, or in other positions including Program Chair of the IEEE NANO 2024 and program Cochair of the 2025 joint Intermag-MMM conference. He is regularly invited to well-established conferences in magnetism and spintronics and he was the organizer of the first international conference on Ising machines.